%% file: main.tex
\documentclass[sigconf]{aamas} 

\usepackage{balance} 

\usepackage{pifont}
\newcommand{\cmark}{\ding{51}}%
\newcommand{\xmark}{\ding{55}}%
\usepackage{babel}
\usepackage{multicol}
\usepackage{subcaption}
\usepackage{amsmath}
\usepackage{enumitem}
\usepackage{algorithm}
\usepackage{algpseudocode}
\DeclareUnicodeCharacter{3BC}{\ensuremath{\mu}}
\setcopyright{ifaamas}
\acmConference[AAMAS '24]{Proc.\@ of the 23rd International Conference
on Autonomous Agents and Multiagent Systems (AAMAS 2024)}{May 6 -- 10, 2024}
{Auckland, New Zealand}{N.~Alechina, V.~Dignum, M.~Dastani, J.S.~Sichman (eds.)}
\copyrightyear{2024}
\acmYear{2024}
\acmDOI{}
\acmPrice{}
\acmISBN{}

\acmSubmissionID{405}

\title[Linking Vision and Multi-Agent Communication through Visible Light Communication using Event Cameras]{Linking Vision and Multi-Agent Communication through Visible Light Communication using Event Cameras}

\author{Haruyuki Nakagawa}
\affiliation{
  \institution{Tokyo Institute of Technology}
  \city{Tokyo}
  \country{Japan}}
\affiliation{  
  \institution{Sony Semiconductor Solutions Corp.}
  \city{Atsugi}
  \country{Japan}}
\email{nakagawa.h.aj@m.titech.ac.jp}

\author{Yoshitaka Miyatani}
\affiliation{
  \institution{Sony Semiconductor Solutions Corp.}
  \city{Atsugi}
  \country{Japan}}
\email{Yoshitaka.Miyatani@sony.com}

\author{Asako Kanezaki}
\affiliation{
  \institution{Tokyo Institute of Technology}
  \city{Tokyo}
  \country{Japan}}
\email{kanezaki@c.titech.ac.jp}

\input{0_abstract}


\newcommand{\BibTeX}{\rm B\kern-.05em{\sc i\kern-.025em b}\kern-.08em\TeX}

\makeatletter
\gdef\@copyrightpermission{
	\begin{minipage}{0.3\columnwidth}
		\href{https://creativecommons.org/licenses/by/4.0/}{\includegraphics[width=0.90\textwidth]{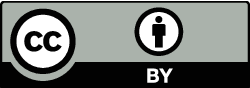}}
	\end{minipage}\hfill
	\begin{minipage}{0.7\columnwidth}
		\href{https://creativecommons.org/licenses/by/4.0/}{This work is licensed under a Creative Commons Attribution International 4.0 License.}
	\end{minipage}
	\vspace{5pt}
}
\makeatother

\begin{document}


\pagestyle{fancy}
\fancyhead{}


\maketitle 


\input{1_introduction}

\input{2_related}
\input{3_approach}
\input{4_simulation}
\input{5_results}

\input{6_conclusion}

\input{8-Acknowledgements}

\newpage


\bibliographystyle{ACM-Reference-Format} 
\bibliography{main}


\input{7_appendix}


\end{document}

%% file: 0_abstract.tex
\begin{abstract}

Various robots, rovers, drones, and other agents of mass-produced products are expected to encounter scenes where they intersect and collaborate in the near future. In such multi-agent systems, individual identification and communication play crucial roles. In this paper, we explore camera-based visible light communication using event cameras to tackle this problem. An event camera captures the events occurring in regions with changes in brightness and can be utilized as a receiver for visible light communication, leveraging its high temporal resolution. Generally, agents with identical appearances in mass-produced products are visually indistinguishable when using conventional CMOS cameras. Therefore, linking visual information with information acquired through conventional radio communication is challenging. We empirically demonstrate the advantages of a visible light communication system employing event cameras and LEDs for visual individual identification over conventional CMOS cameras with ArUco marker recognition. In the simulation, we also verified scenarios where our event camera-based visible light communication outperforms conventional radio communication in situations with visually indistinguishable multi-agents. Finally, our newly implemented multi-agent system verifies its functionality through physical robot experiments.

\end{abstract}

%% file: 1_introduction.tex
\section{Introduction}

Multi-agent systems are becoming increasingly important in various fields~\cite{saeid2020, ismail2018survey,zhou2022swarm}. 
Coordinating agents such as rovers in homes and warehouses and drones for tasks such as exterior inspections, security, and 3D scanning is becoming increasingly necessary. The cooperation of multiple agents can enable efficient task accomplishment in various domains.

\begin{figure}[]
  \centering
  \includegraphics[width=1.00\linewidth]{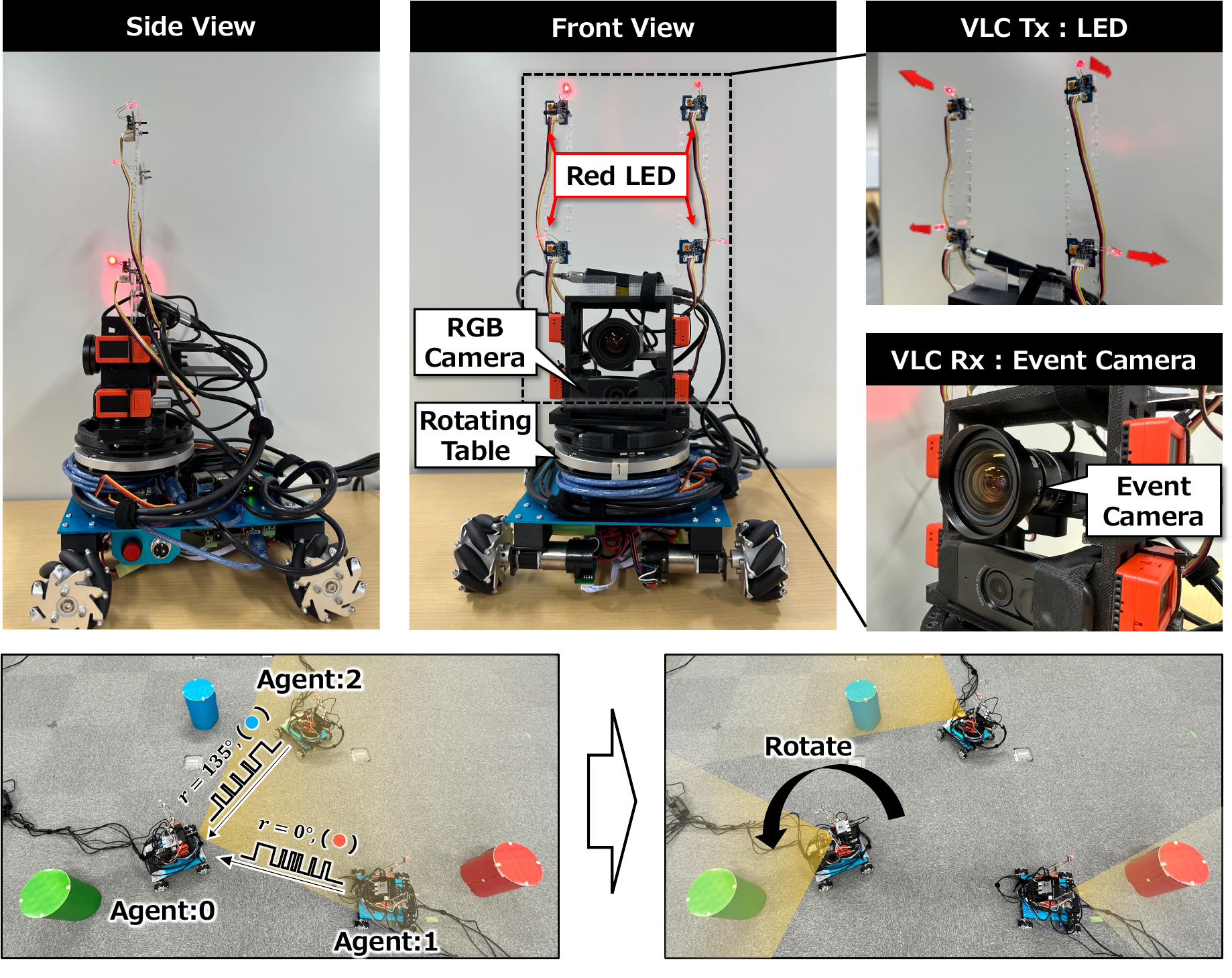}
  \caption{Physical agents used in the multi-agent system with visible light communication using event cameras. Each agent is equipped with an event camera and four separate LEDs facing different directions to increase transmission probability to the surroundings. Each agent can rotate and engage in cooperative behavior by identifying visually identical agents for information sharing.
  }
  \label{fig:zikkiconcept}
\end{figure}

\begin{figure*}[t]
  \centering
  \begin{minipage}{0.31\hsize}
    \centering
    \includegraphics[width=\linewidth]{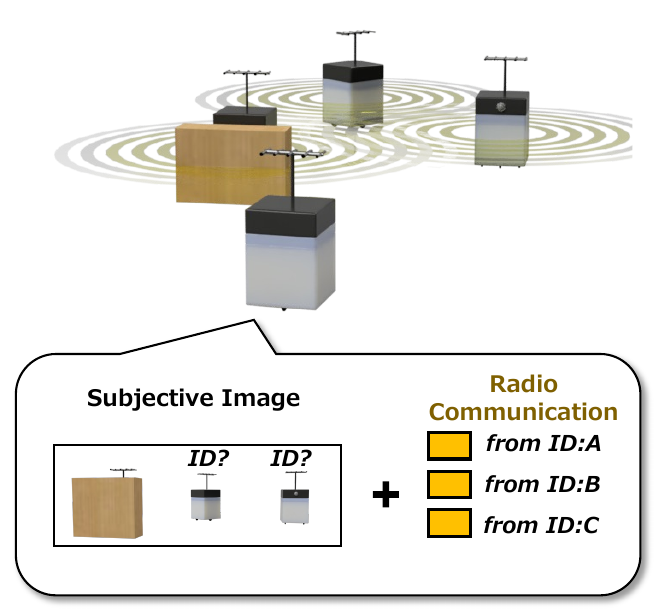}
    \subcaption{Radio communication}
  \end{minipage}
  \begin{minipage}{0.31\hsize}
    \centering
    \includegraphics[width=\linewidth]{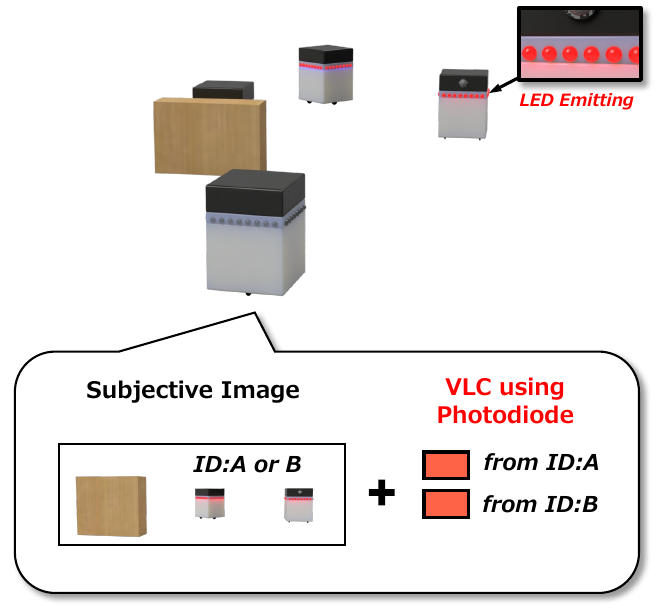}
    \subcaption{Single PD-VLC}
  \end{minipage}
  \begin{minipage}{0.31\hsize}
    \centering
    \includegraphics[width=\linewidth]{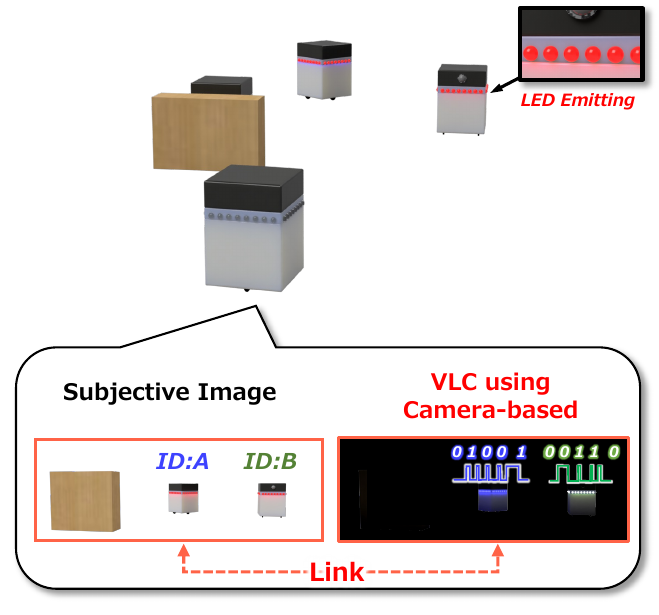}
    \subcaption{RGB-VLC and event-VLC}
  \end{minipage}
  \caption{
    \textcolor{black}{Comparison of communication methods in multi-agent system of robots with identical appearance. (a) Radio wave communication can retrieve communication information from agents beyond a wall. However, it cannot distinguish between two agents captured by visual information from RGB cameras. With VLC-based methods (b) and (c), assuming the RGB camera and receiver for VLC have the same FOV, it is possible to distinguish which agent with a specific ID is present within FOV. However, (b) when using a single photodiode without spatial resolution in the light-receiving unit, it is impossible to determine the identities of the two agents. (c) RGB-VLC and event-VLC can acquire information linked to the spatial location in the image.}
  }
  \label{fig:linkingvisualcomm}
\end{figure*}

Individual identification and communication are two essential technologies in multi-agent systems involving mobile entities. Regarding individual identification, commonly employed methods rely on visual cues such as color and Augmented Reality University of Cordoba (ArUco) markers~\cite{garrido2014automatic, evsvlc2022iros, wu2021spatial}. However, altering their appearance with different colors can be challenging for mass-produced robots or vehicles. ArUco markers pose challenges for moving and rotating objects, requiring multiple large markers for stable recognition from all angles, especially when viewed from a distant 3D perspective, such as with drones.
Communication is crucial in achieving global optimization by allowing agents to acquire information about other agents that operate in a distributed manner~\cite{foerster2016learning,lowe2017multi}. However, in environments where individual identification cannot be achieved, it becomes challenging to associate visual information and signals obtained through radio-based communication (Fig.~\ref{fig:linkingvisualcomm} (a)).


To address these challenges, the present paper explores a multi-agent system equipped with optical wireless communication using event cameras (Fig.~\ref{fig:zikkiconcept}). Optical wireless communication, which is distinct from conventional radio wave communication, is a communication method that utilizes light. When visible light is used, it is referred to as visible light communication (VLC)~\cite{Nasir2019, yamazato2014, evsvlc2020, evsvlc2022iros, Rehman2019}. VLC employs light-emitting devices such as LEDs as transmitters (Tx) and light-receiving devices like photodiodes or cameras as receivers (Rx). One advantage of VLC is its suitability for visual individual identification. Flashing LEDs in devices enables the emission of variable IDs and signals, allowing for decoding at a distance. Compact devices such as LEDs impose fewer constraints on placement, enabling recognition from various angles.

In recent years, there has been significant development in the field of event cameras~\cite{suh20201280, Finateu2020}. An event camera detects temporal contrasts that exceed a predefined threshold on a per-pixel basis and outputs them as ``events.'' Hence, it operates differently than conventional CMOS cameras, offering high temporal resolution and sparse data acquisition. Such event cameras can capture the rapid flashing of LEDs as changes in brightness, serving as a receiver for VLC~\cite{evsvlc2022iros, Chen2020, Tang2022}. 

Table \ref{comparevlc} shows a summary of the findings from the literature~\cite{Rehman2019,shen2018vehicular,evsvlc2022iros} on comparing VLC methods. We refer to the method using a photodiode as \textit{PD-VLC}, the method using a conventional RGB camera as \textit{RGB-VLC}, and the method utilizing an event camera as \textit{Event-VLC}. 
First, \textit{PD-VLC} decodes optical signals at high frequencies from megahertz to gigahertz, enabling faster communication speeds~\cite{viola201715}.
However, a single optical receiver is susceptible to interference from ambient light and different signal sources, making it challenging to separate multiple signals in multi-agent scenarios~\cite{shen2018vehicular, Rehman2019} (Fig.~\ref{fig:linkingvisualcomm} (b)).
On the other hand, the data rate of \textit{RGB-VLC} is very low because it can generally capture images only at several tens of frames per second (FPS). 
\textit{Event-VLC} can operate tens of times faster than a conventional CMOS camera and produce a relatively high data rate. Notably, a photodiode has a limited communication range, while camera-based methods such as \textit{RGB-VLC} and \textit{Event-VLC} can be broader. This implies that a photodiode cannot widen the Field of View (FOV) to ensure reliable communication with Tx while avoiding ambient light and other signal sources. These are the advantages of camera-based schemes in multi-agent systems regarding signal separation availability (Fig.~\ref{fig:linkingvisualcomm} (c)). Regarding whether decoding is possible robustly,  the smaller dynamic range of the \textit{RGB-VLC} compared with the \textit{Event-VLC} makes it challenging to communicate. 

\begin{table}[t]
\centering
\caption{\textcolor{black}{Comparison of VLC method characteristics}}
\label{comparevlc}
\begin{tabular}{@{}lccc@{}}
\toprule
                & \textit{PD-VLC} & \textit{RGB-VLC} & \textit{Event-VLC}   \\ \midrule
Data Rate       & \cmark~\textbf{High}   & \xmark~Low                 & \cmark~Mid                \\
Source Separate & \xmark~No              & \cmark~\textbf{Yes}        & \cmark~\textbf{Yes}       \\
Communication Range           & \xmark~Narrow          & \cmark~\textbf{Wide}       & \cmark~\textbf{Wide}      \\
Decode Robustness      & \cmark~Mid             & \xmark~Low                 & \cmark~\textbf{High}      \\ \bottomrule
\end{tabular}
\end{table}

The contributions of this paper are as follows:
\begin{itemize}[leftmargin=*]
\item We demonstrated the utility of an event camera-based VLC system for visually identifying moving multi-agent entities.
\item Through simulation experiments, we demonstrated that a camera-based VLC system can perform visually indistinguishable multi-agent tasks better than a conventional system using an RGB camera and radio communication.
\item We constructed physical robots equipped with event cameras and conducted real-world experiments, demonstrating the practicality of a multi-agent system based on event-VLC for the first time.
\end{itemize}

%% file: 2_related.tex
\section{Related Work}

\noindent \textbf{Multi-agent system with communication.} In multi-agent environments, numerous reports have shown the benefits of utilizing communication among agents to enhance task accomplishment while promoting behavioral diversity~\cite{foerster2016learning,lowe2017multi}. Additionally, in multi-agent systems, as the number of agents increases, the programming of behaviors becomes increasingly complex. Therefore, many proposals have suggested using multi-agent reinforcement learning to automatically learn policies by setting rewards.
Forester et al.~\cite{foerster2016learning} introduced \textcolor{black}{Differentiable Inter-Agent Learning (DIAL)} in deep recurrent Q-networks, demonstrating high performance using end-to-end differentiable communication channels among different agents. Furthermore, Lowe et al.~\cite{lowe2017multi} proposed the \textcolor{black}{Multi-Agent Deep Deterministic Policy Gradient (MADDPG)} algorithm, showcasing the ability to learn advanced cooperative behaviors, even in scenarios involving adversarial tasks and communication.

Some research papers have taken into account realistic environment constraints, such as communication range, error rates, and budgets, for radio communication~\cite{wang2019r, Agarwal-2019-112867,atoc2018}. Furthermore, in ideal simulation environments, precise location information is assumed to be obtainable for each agent. However, in practical scenarios, such as urban settings with buildings, underwater environments, or environments with significant radio noise, it is known that location information cannot be obtained or its accuracy is degraded~\cite{ross2019augmenting,millard2019map}. Additionally, many existing multi-agent reinforcement learning systems, whether using state information such as position and velocity or visual information, often assume that agents can distinguish among one another~\cite{lowe2017multi,wang2019r}. These assumptions may not be applicable when dealing with mass-produced items that look identical or when agents are too far away or move too quickly for reliable recognition. 
We explore the application of VLC in multi-agent environments as a solution to these challenges.

\noindent \textbf{Visible light communication (VLC).} VLC is a method of communication that uses light.~\cite{Rehman2019, SHAABAN2021483}. Because of its use of light, VLC offers several advantages, such as the ability to incorporate communication functionality into existing devices like displays, LEDs, and cameras and the potential for secure communication that does not leak information to unintended recipients. Additionally, VLC has garnered attention for its applicability in environments with high radio frequency noise, places where radio waves are undesirable (e.g., hospitals, factories), and conditions where radio waves are absorbed, such as underwater communication~\cite{Tiansi2010, Nasir2019}.

The camera-based approach utilizes a two-dimensional photodiode array with the image sensor of a typical camera for VLC. This approach has the advantage of signal separation when multiple light sources are captured in the camera's field of view. 
However, conventional CMOS camera-based systems suffer from lower data rates than single photodiode-based systems, which are typically operating at around 30 to 60 frames per second (fps).
The VLC system employed in the current paper, which uses an event camera, leverages the benefits of the camera-based approach while achieving higher data rates through a high frame rate. Research on the use of event cameras for VLC has gained momentum in recent years, especially in automotive applications~\cite{shen2018vehicular,evsvlc2022iros}.

Although not visible light, studies on multi-agent systems using optical wireless communication have been reported~\cite{nakagawa2023}. However, the limited communication range due to the use of a photodiode as the receiver prevents full operation in physical multi-agent environments.

\noindent \textbf{Event cameras.} An event camera is a type of sensor that detects changes in luminance, which is significantly different from a conventional CMOS camera that acquires absolute luminance values on a frame-by-frame basis~\cite{suh20201280, Finateu2020,gallego2020event,scheerlinck2019ced}. Event cameras can capture changes in log luminance from their circuit characteristics, allowing for a wide dynamic range, and they output luminance changes asynchronously as ``events'' at microsecond-level speeds. Applications using event cameras have been reported in various fields, including computer vision, robotics, and VLC~\cite{vitale2021event,vemprala2021representation,kim2008simultaneous,pan2020single}.

In recent years, RGB hybrid event-based vision sensors that combine RGB pixels and event pixels have also emerged~\cite{kodama2023,Guo2023hybrid}. These sensors can acquire RGB information, in addition to high-speed event information, leading to investigations into using event information for image quality correction of RGB data~\cite{xu2021motion,zhang2020hybrid}. 
These hybrid sensors are valuable for tasks such as image quality enhancement and applications that leverage both RGB observations and event information.

\noindent \textbf{Identification.} 
There are various methods for identifying individual agents in multi-agent systems. One approach is visual, where methods such as changing the agents' color or using identification tags like ArUco markers are employed~\cite{evsvlc2022iros, wu2021spatial}. However, ArUco markers lose their robustness against motion blur and defocus when attached to moving objects~\cite{toyoura2014mono, cejka2018improving}. 
There are other methods for ID reading using radio frequency (RF), such as the one reviewed in~\cite{zhang2017review}, and approaches like ultra wide band (UWB)~\cite{Minoli2018, Coppens2022} that enable spatial positioning. 
However, it is essential to note that while increasing radio wave intensity allows for long-distance communication, there are often regulatory constraints on power levels and limits on the number of simultaneous communications. As a result, it can be challenging to determine the direction of target agents located more than several tens of meters away.



%% file: 3_approach.tex
\section{Multi-agent system with event-VLC} \label {approach}

\begin{figure*}[t]
  \centering
  \begin{minipage}{0.32\hsize}
    \centering
    \includegraphics[width=\linewidth]{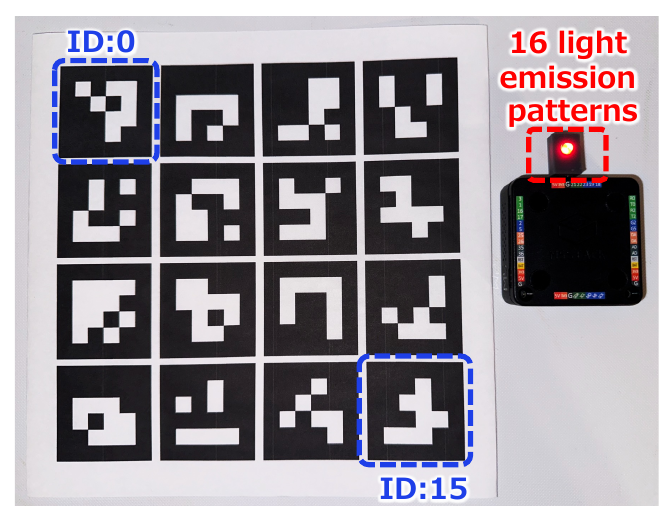}
    \subcaption{ArUco markers and active LED markers}
  \end{minipage}
  \begin{minipage}{0.34\hsize}
    \centering
    \includegraphics[width=.94\linewidth]{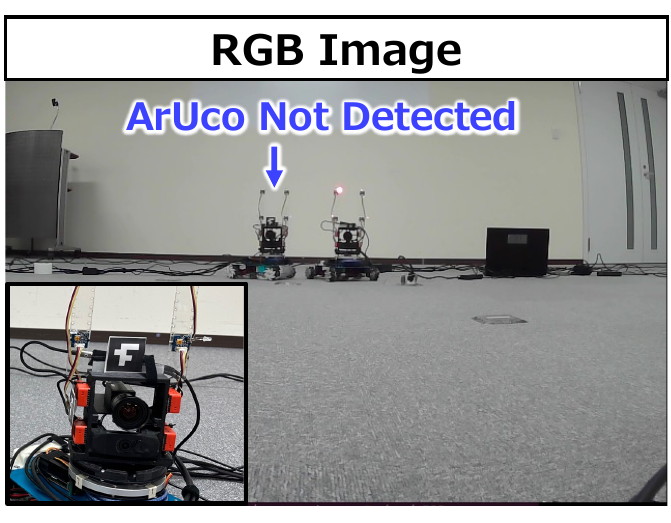}
    \subcaption{ArUco marker detection using an RGB camera}
  \end{minipage}
  \begin{minipage}{0.32\hsize}
    \centering
    \includegraphics[width=\linewidth]{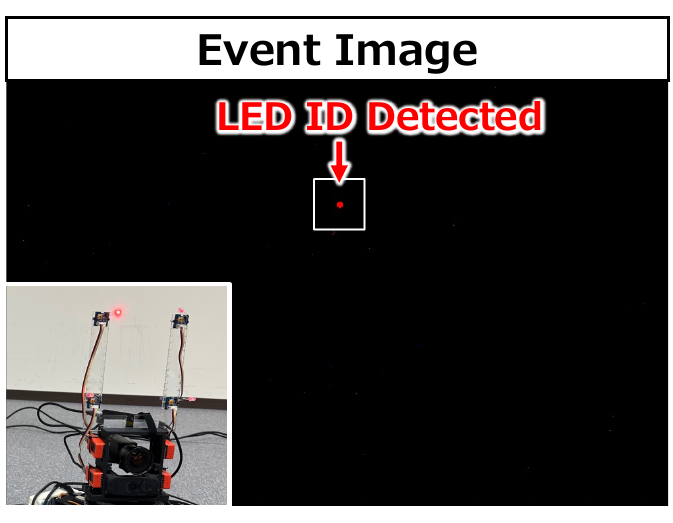}
    \subcaption{LED detection using an event camera}
  \end{minipage}
  \caption{
    \textcolor{black}{(a) ArUco markers and active LED markers for visual identification, (b) ArUco marker detection example, and (c) LED detection example. Because of the blurred resolution and focus, ArUco marker detection failed, whereas active LED markers were successfully detected using an event camera.}
  }
  \label{fig:ledarucomarker}
\end{figure*}

\subsection{Design}
\label {approach_actualrobotsingle}
Incorporating the event-VLC system into a multi-agent system is advantageous for individual identification and communication. The setup considered for practical implementation is illustrated in Fig.~\ref{fig:zikkiconcept}. The event camera necessary for VLC operation was installed coaxially with an RGB camera, with approximately the same horizontal field of view. 
For the RGB cameras, we used Buffalo BSW500MBK 2 megapixel web camera in this paper.
For the event camera, we used a camera with 0.92 megapixel event-based vision sensor IMX636.
We basically applied the method~\cite{censi2013low} to the modulation and decoding process of LED signals.
Four LEDs as the Tx were attached to acrylic boards and installed. At this point, the four LEDs were oriented in separate directions, each at a 90$^{\circ}$ angle in the horizontal plane. 
Each LED was commanded with transmission signals from M5stick devices, and these M5stick devices were connected to Jetson Nano via serial communication. All these components were mounted on a rover equipped with a rotating platform, enabling the measurement of rotation angles and distances traveled through servo motors. The rotary table can operate in the range of 0$^{\circ}$ to 180$^{\circ}$. Therefore, the information acquired through RGB and event cameras and self-rotation angle information is encoded and transmitted through LED blinking. 






\begin{table}[]
\centering
\caption{\textcolor{black}{Performance comparison for identification}}
\label{compareid}
\begin{tabular}{@{}lcc@{}}
\toprule
                      & \hspace{-3mm} RGB w/ ArUco & \ Event-VLC w/ LED                   \\ \midrule
Distance Limit        & 2 m       & \textbf{5.5 m}                \\
Speed Limit        & 50 cm/sec & >\textbf{100 cm/sec}             \\
Brightness Robustness & Low       & \textbf{High}                    \\
Obscured Limit        & 20 \%     & \textbf{\textgreater{}90\%}   \\
Signal Mutability            & Fixed     & \textbf{Changeable}          \\ \bottomrule
\end{tabular}
\end{table}


%
%


\subsection{Comparison with other visual identification methods} \label {compareidmethod}
\begin{figure*}[t]
  \centering
  \begin{minipage}{0.31\hsize}
    \centering
    \includegraphics[width=\linewidth]{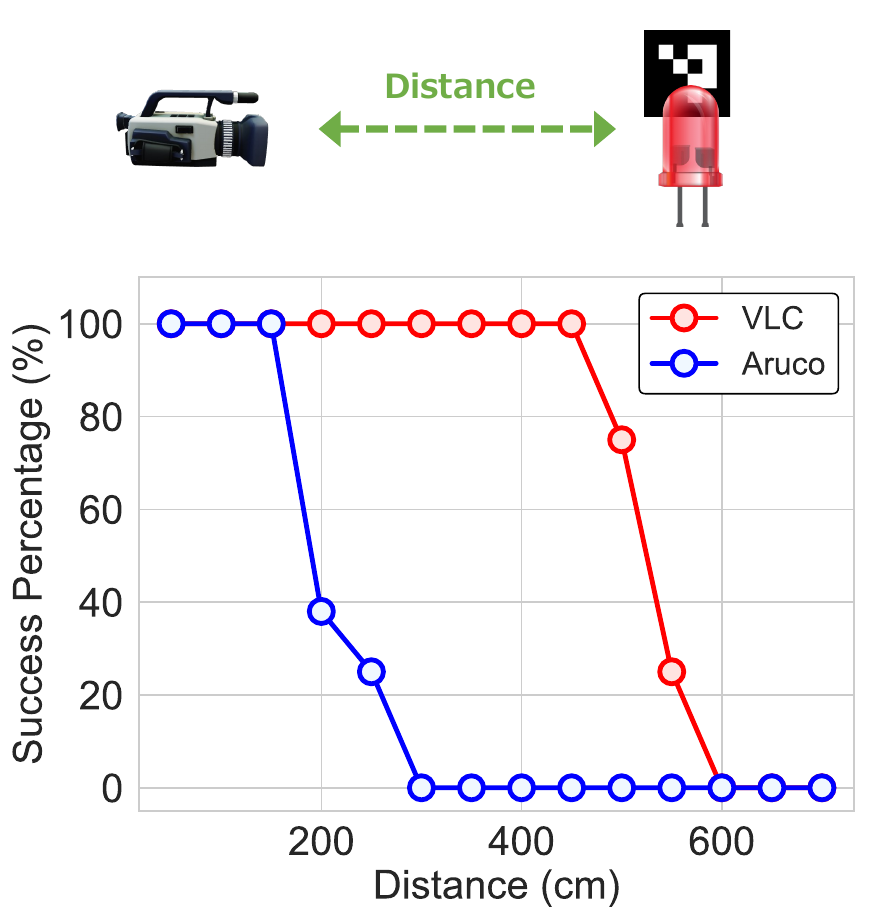}
    \subcaption{Distance dependance}
  \end{minipage}
  \begin{minipage}{0.31\hsize}
    \centering
    \includegraphics[width=\linewidth]{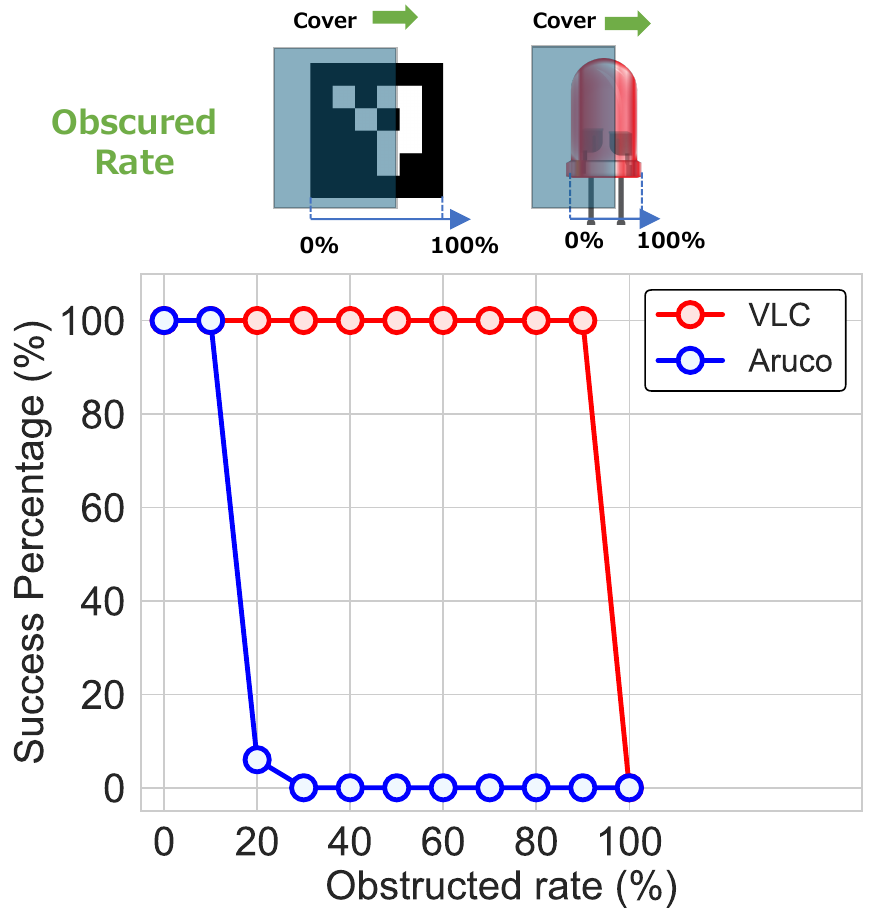}
    \subcaption{Obscured rate dependance}
  \end{minipage}
  \begin{minipage}{0.31\hsize}
    \centering
    \includegraphics[width=\linewidth]{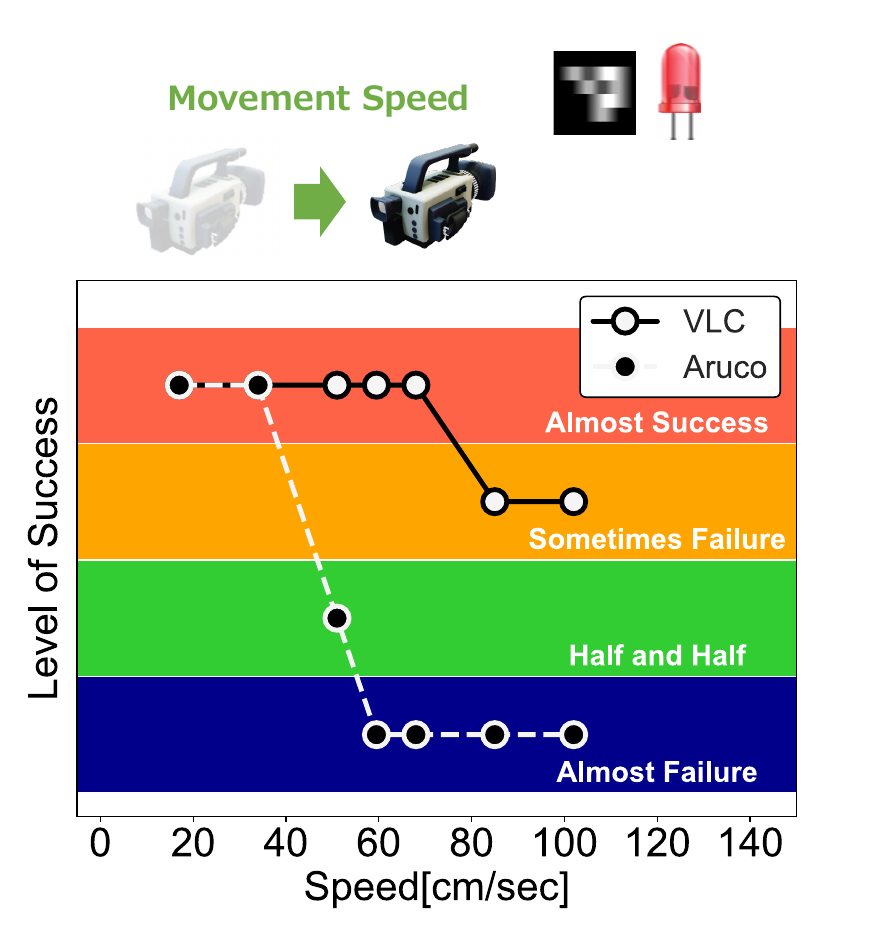}
    \subcaption{Movement speed dependance}
  \end{minipage}
  \caption{
    \textcolor{black}{Comparison of recognition accuracy of ArUco markers using RGB cameras and LED active markers using event cameras. (a) The event-VLC method enables communication over longer distances. Event-VLC is also successful (b) when the markers are partially obstructed and (c) when blurring occurs because of movement, compared with the case with RGB cameras.}
  }
  \label{fig:comparisonrecog}
\end{figure*}

In this section, we compared the performance of visual ID that is crucial in multi-agent systems, using \textit{Event-VLC} and ArUco marker recognition using an RGB camera.
Table \ref{compareid} and Fig.~\ref{fig:comparisonrecog} present the comparison results, specifically examining their suitability for integration into multi-agent systems, which include mobile entities.

For the ArUco markers, we assumed a size of approximately 4 cm on each side, which is similar to the dimensions of the modules capable of driving LEDs (Fig.~\ref{fig:ledarucomarker} (a)). ArUco markers were detected using modules within OpenCV. 
For RGB cameras, Buffalo BSW500MBK 2 megapixel web cameras were primarily used, while we used FLIR Blackfly S USB 3.1 camera for the experiments to verify speed dependency. 
Red LEDs were employed as the Tx for the event-VLC. The flashing period of the LED was set to 4 milliseconds.
Throughout the experiments, we maintained fixed focus and imaging conditions. Regarding recognition rate, we calculated the changes in distance and occlusion based on how many of the 16 IDs could be recognized. Additionally, to assess resistance to speed, we plotted the degree of recognition rate from the results obtained by moving the camera attached to the robot arm to recognize a ID and then converting it to 30 frames per second equivalent, categorizing it into four levels of recognition rate.

First, regarding the extent to which recognition is possible even at a distance, it was found that \textit{Event-VLC} can recognize objects at a greater distance than \textit{RGB w/ ArUco} (Fig.~\ref{fig:comparisonrecog}(a)). Although the recognition accuracy of ArUco markers depends on resolution and focus, \textit{Event-VLC}, in principle, can recognize objects as long as there is a response in at least one pixel. 

Next, we investigated the impact on recognition rates when obstacles partially cover the target markers. For ArUco markers, recognition becomes impossible when they are covered by approximately 20\%. In contrast, recognition rates do not significantly decrease for LEDs, even when covered by up to approximately 90\% (Fig.~\ref{fig:comparisonrecog}(b)). This is an essential characteristic in situations where agents themselves can become obstacles in multi-agent scenarios.

Finally, regarding robustness against motion blur, \textit{Event-VLC} demonstrates the ability to maintain recognition at high speeds because of its high temporal resolution and tracking technology (Fig.~\ref{fig:comparisonrecog}(c)). On the other hand, it was found that typical 30 FPS cameras, such as RGB Camera, cannot achieve recognition beyond a certain speed because of the impact of motion blur. This is believed to be because ArUco markers rely on recognizing high-frequency patterns and are susceptible to motion blur effects. 
Furthermore, \textit{Event-VLC} exhibits superior robustness to ambient lighting conditions. This is attributed to the advantage of the LED-based illumination system and the high dynamic range of \textit{Event-VLC}. To reduce motion blur in RGB cameras by increasing the FPS, it is essential to strike an optimal balance because a decrease in exposure time can lead to darker images and a reduction in the recognition rate. Also, \textit{Event-VLC} offers the advantage of changing the signal later. 
These results suggest the utility of event-VLC in a wide range of environments, including outdoor settings, within multi-agent systems.
As shown in Fig.~\ref{fig:ledarucomarker} (b) and (c), when comparing the setup with ArUco markers on the physical device and VLC using LEDs, ArUco markers failed to be recognized at certain distances, whereas LED signals could be detected using event cameras.

%% file: 4_simulation.tex
\section{simulation experiments}\label {4_sim}

\subsection{Settings} \label {simsetting}
Based on the analysis in Section \ref{approach}, it is plausible that there are scenarios where individual identification using RGB cameras is unattainable. As a result, we proceeded with the assumption that individual identification would not be feasible in the given scenario. Also, the location information was set to be unobtainable.
In the present study, we have established the conditions in a two-dimensional plane using the Multi-Agent Particle Environments (MAPE)~\cite{lowe2017multi} environment, where one-dimensional visual information can be obtained (Appendix~\ref{appendix_1dobs}).
In the VLC configuration, communication info was assumed to come only from visible agents. Obstructions by agents or landmarks prevent signal acquisition in this setup. The communication method used in the present study with event-VLC provides communication information linked to visual information. Therefore, in the case of the event-VLC, individual identification is assumed to be possible.

\begin{figure*}[t]
  \centering
  \includegraphics[width=0.85\linewidth]{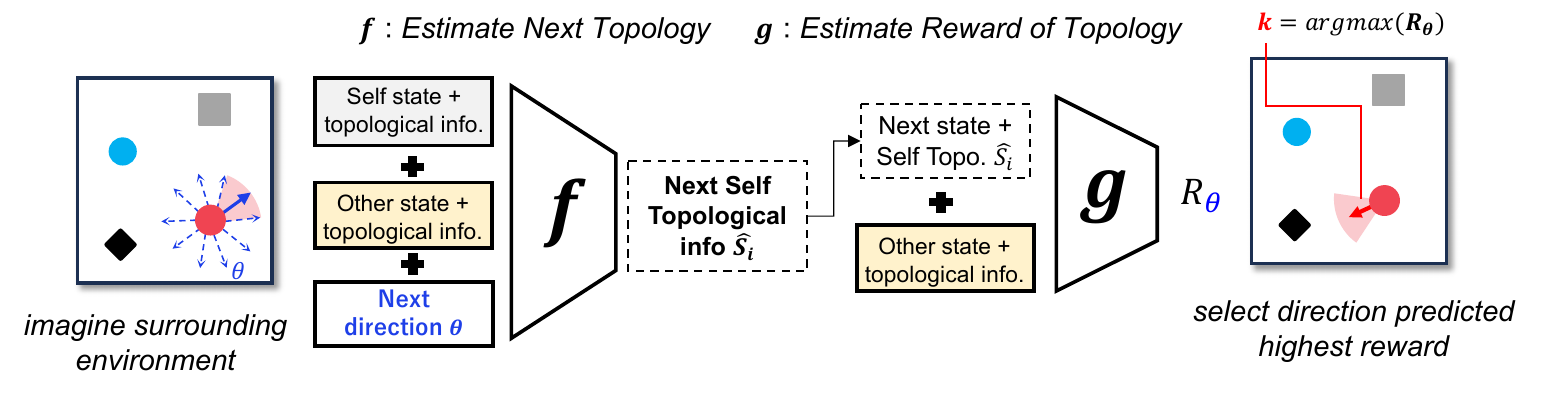}
  \caption{\textcolor{black}{For cooperative actions using limited vision, such as \textit{Simple Spread} and \textit{Predator-Prey} tasks, agents imagine what they would see if they changed their visual direction and predict whether they would earn a higher reward. }
  }
  \label{fig:model_for_sspp}
\end{figure*}

\begin{table}[t]
\centering
\caption{Task settings for the simulation experiments}
\label{taskbasic}

\begin{tabular}{@{}lp{2cm}p{2cm}l@{}}
\toprule
Task & Communication & Action & FOV \\
\midrule
\textit{\textbf{Simple Spread}} & view direction, visible objects & view direction, movement [m] & 120\textdegree \\
\textit{\textbf{Predator-Prey}} & view direction, visible objects & view direction, movement [m] & 120\textdegree \\
\textit{\textbf{Simple Swing}} & view direction, visible objects & view direction & 80\textdegree \\
\textit{\textbf{Target Encirclement}} & velocity & Force [N] & 360\textdegree \\
\textit{\textbf{Goal Crossing}} & velocity, vector to goal & Force [N] & 360\textdegree \\
\bottomrule
\end{tabular}
\end{table}

\subsection{Tasks}

We present basic information regarding the tasks employed for simulation in Table \ref{taskbasic}. Further information is provided below.

\noindent \textbf{$\blacktriangleright$Simple Spread} To assess the cooperation of agents with limited vision in a task involving movement, we employed the \textit{Simple Spread} task. In this task, there are three agents and three landmarks, and each agent learns to reach a different landmark. The reward structure is similar to the standard setup, where a penalty is incurred based on the total distance to the nearest agent for each landmark.

\noindent \textbf{$\blacktriangleright$Predator-Prey} In this task, there are three good agents and one adversarial agent. The adversarial agent's objective is to evade the good agents, while the good agents aim to capture the adversarial agent. The reward structure is consistent with the standard setup, where penalties are assigned based on the distance to the adversarial agent, and rewards are granted when the good agents successfully capture the adversarial agent. Note that the good agent is placed randomly in the field with respect to its initial position, while the adversarial agent is placed in the center of the field. Therefore, the adversarial agent should learn the policy of escaping to the outside. In the benchmarking, the same learned policies were applied as adversarial agents, regardless of the communication method.

\noindent \textbf{$\blacktriangleright$Simple Swing} We designed an original task called \textit{Simple Swing} that considers experiments conducted with physical agents. In this task, there are three agents and three landmarks, and 
agents must change their visual orientation to avoid losing sight of the landmark. As a reward, a penalty is assigned for the number of missed landmarks. Furthermore, in the case of VLC, rewards were provided when communication was established. 
This task requires agents to better understand their surroundings beyond their individual information by collaborating with other agents with limited visual information. As shown in Table \ref{taskbasic}, each agent communicates information about its visual direction and the entities it can see. The initial positions of agents are randomly placed along the edges of an equilateral triangle. The maximum angle that can be rotated at one time is 30$^{\circ}$.

\noindent \textbf{$\blacktriangleright$Target Encirclement} This is a task for agents to perform adjustments dynamically while interacting with neighboring agents. In this task, five agents need to gather around a target object while maintaining an appropriate distance in the shortest possible time. During this task, each agent must act to avoid collisions with other agents and the target. As the reward setup, the total sum of distances to the targets for each agent and the number of collisions is penalized. In this scenario, the agent is set to move according to the equations of motion and hence inertia must be taken into account in training.

\noindent \textbf{$\blacktriangleright$Goal Crossing} In this task, each agent is assigned a different goal position. The agents learn to reach their respective goals in the shortest possible time while avoiding collisions with other agents. Agents do not have access to their own positions, but they can obtain the vectors to their goal position. As for the reward configuration, it was set such that the total distance to each agent's respective goal and the number of collisions were penalized. 

\subsection{Models and Algorithms}

\begin{figure*}[t]
  \centering
  \begin{minipage}{0.19\hsize}
    \centering
    \includegraphics[width=\linewidth]{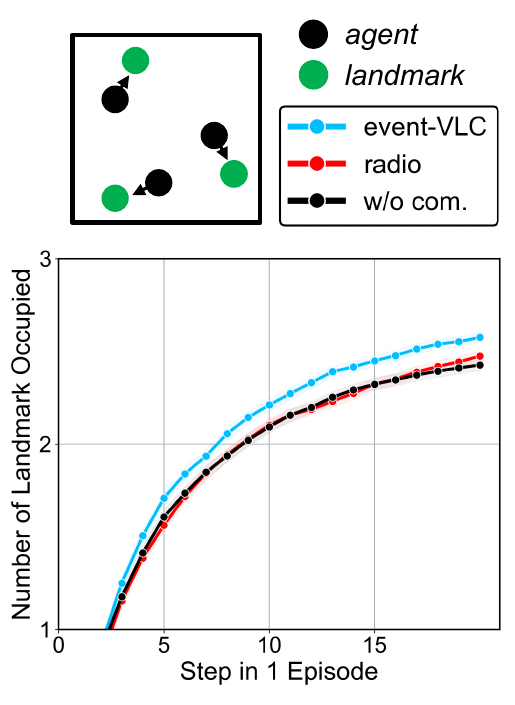}
    \subcaption{\textit{Simple Spread}}
  \end{minipage}
  \begin{minipage}{0.19\hsize}
    \centering
    \includegraphics[width=\linewidth]{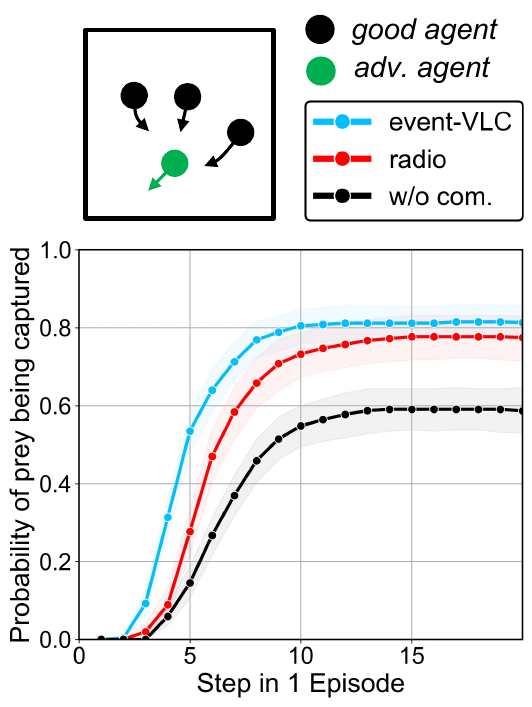}
    \subcaption{\textit{Predator-Prey}}
  \end{minipage}
  \begin{minipage}{0.19\hsize}
    \centering
    \includegraphics[width=\linewidth]{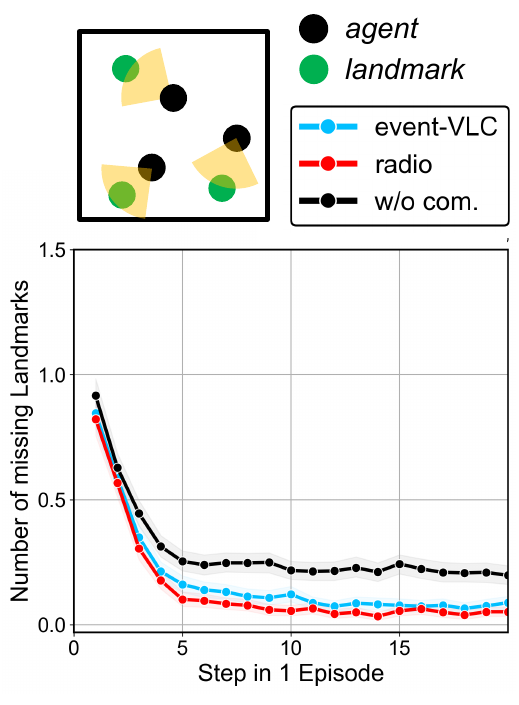}
    \subcaption{\textit{Simple Swing}}
  \end{minipage}
  \begin{minipage}{0.19\hsize}
    \centering
    \includegraphics[width=\linewidth]{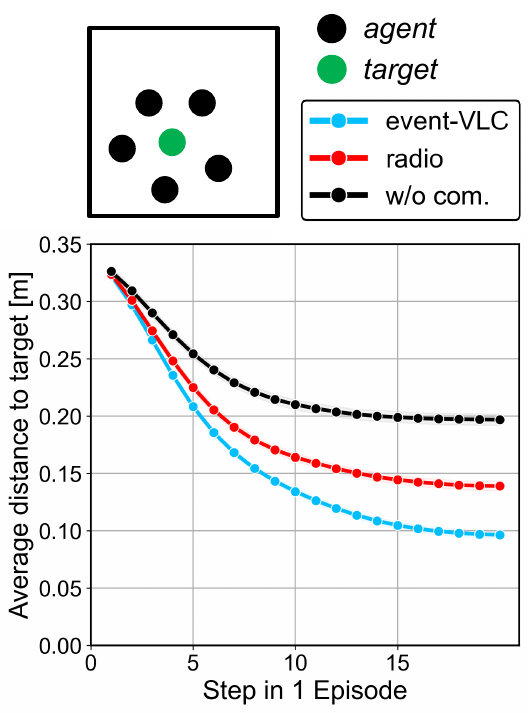}
    \subcaption{\textit{Target Encirclement}}
  \end{minipage}
  \begin{minipage}{0.19\hsize}
    \centering
    \includegraphics[width=\linewidth]{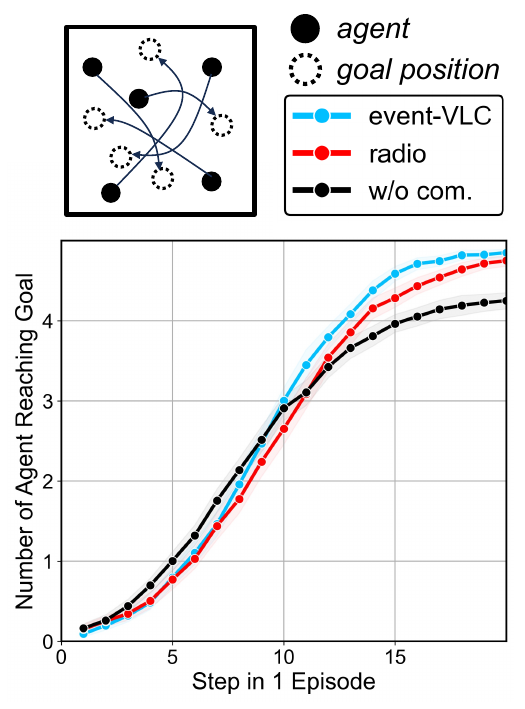}
    \subcaption{\textit{Goal Crossing}}
  \end{minipage}
  
  \caption{
    Simulation benchmark results for each task per 1 episode (= 20 steps). In scenarios where visually identifying other agents proves difficult, camera-based methods such as event-VLC have shown superior accuracy.
  }
  \label{fig:simulationresult}
\end{figure*}

\noindent \textbf{$\blacktriangleright$Simple Spread, Predator-Prey} First, for the tasks of \textit{Simple Spread}, \textit{Predator-prey}, we employed a reinforcement learning algorithm that incorporates limited vision~\cite{nakagawa2023}. As illustrated in Fig.~\ref{fig:model_for_sspp}, agents learn visual direction changes separately from movement actions. In terms of learning actions, reinforcement learning is conducted through the MADDPG~\cite{lowe2017multi} algorithm to determine how much to advance or when to stop. The models utilized for learning behaviors involve a one-dimensional convolutional layer shared between the actor and critic networks, processing one-dimensional visual data (Appendix~\ref{appendix_1dobs}).

We employed a network $f$ that learns what might be visible when altering the visual direction. In this process, each agent maintains ``topological information'', a bit sequence about whether entities are visible to them. Agents capable of communication can also retrieve topological information from other agents. Concatenating this information with state data gathered over the past three frames, communication data from other agents are simultaneously acquired and stored as $V_i^{t}$.

We feed this combined information and imagined visual direction $\theta$ into the network $f$ to predict self topological information $\hat{S}_i^{t+1}$ that will be visible when facing that angle $\theta$, as in the following formula:
%
\begin{equation}
  \hat{S}_i^{t+1} = f\left(\{\{\mathbf{d}_i^{t'}\}_{i\in a_c}, V_i^{t'}\}_{t'=t-T+1}^t,\theta\right).
  \label{eq:f}
\end{equation}
In this context, $a_c$ denotes the collective of agents, including $a_i$ itself, with whom communication is possible. In addition, $d_i$ is looking direction of $a_i$. The visual direction of $a_c$ is also incorporated as input variables. 
Furthermore, we employ a network $g$ to learn when altering the visual direction would likely yield higher rewards than the current state. This network was trained as follows:
\begin{equation}
  P(r_i^{t+1} - r_i^t > \Delta r) = g\left(\{\mathbf{d}_i^{t'}\}_{i\in a_c}, \hat{V}_i^{t'}\}_{t'=t-T+1}^t\right).
  \label{eq:g}
\end{equation}
At execution, each agent predicts what is visible in every $2\pi/N_D$ direction using $f$ and selects the direction with the highest expected reward from predictions made by $g$. In this paper, $N_D=36$ and $T=3$ were set. The parameter $\Delta r$ was set to 0.001.

\noindent \textbf{$\blacktriangleright$Simple Swing} For the \textit{Simple Swing} task, which does not involve movement, we partially employed the methodology mentioned above and adapted it for experimental conditions. 
Each agent retains information about its visual direction and the entities it can see based on visual data. 
Within the critic model, individual observations and actions are combined, and network $f$ is trained to predict the next self topological information ${S}_i^{t+1}$ after altering the visual direction by actions. Details of the model are included in the Appendix~\ref{appendix_simpleswingmodel}.
The loss function of predicting the next topological information is represented by the following equation:
\begin{equation}
\textcolor{black}{
  \mathcal{L}_N = \frac{1}{N} \sum_{i=1}^N \| S_i^{t+1} - \hat{S}_i^{t+1} \|^2. \label{eq:L_n}
}
\end{equation}
Subsequently, the concatenated vector is processed by the MLP network $C$ to output Q values. Learning using TD error loss $\mathcal{L}_Q$ is backpropagated only through the network $C$. Critic's loss function in the reinforcement learning algorithm used in this study is described in Appendix~\ref{appendix_preliminary}. Actor is trained solely through MLP from their individual observations.
In this task, simplified information about what is visible, rather than high-dimensional visual data, is utilized as observations. This approach aims to conduct physical experiments without considering the Sim2Real aspects.


\noindent \textbf{$\blacktriangleright$Target Encirclement, Goal Crossing} For these tasks, we utilized a model outputting forces [N] in the X and Y directions through shared 1-dimensional convolutional neural network (CNN) for critic and actor because the agents have omnidirectional vision. In this case, for \textit{Event-VLC}, we adopted a model that combines communication signals in the channel direction of visual information input to CNN, allowing communication and visual processing to pass through the same flow.

\subsection{Results}

The results of 10K times benchmark on the trained model are shown in Fig.~\ref{fig:simulationresult}. The trajectory of the agent in tasks requiring movement is depicted in Fig.~\ref{fig:trajimage} of the Appendix~\ref{appendix_radioid}.
First, for the \textit{Simple Spread} task, the performance of the \textit{Event-VLC} system was superior. This can be attributed to the ability to comprehend the surroundings by linking which agent conveys communication information when coordinating limited visual information in the case of event-VLC. 
This trend was similarly observed in the \textit{Predator-prey} task. The adversarial agent endeavors to escape toward the outer boundaries by maneuvering through the spaces between the good agents. However, the good agents collaborate visually, enabling them to pinpoint the position of the adversarial agent swiftly.


Next, concerning the \textit{Simple Swing} task, the performance of \textit{Event-VLC} was comparable to that of radio communication. Despite the disadvantageous setting for VLC, where a limited field of view makes communication with other agents difficult, the ability of event-VLC to associate location and visual information in communication compensates for this, resulting in comparable performance to radio communication, which can obtain all communication information. When training with the ability to predict the next visual input, as utilized in the model architecture in the present paper, the performance surpasses simple training through MLP. Furthermore, the configuration of providing rewards for VLC communication itself also contributes to the overall performance improvement. More details are provided in Appendix~\ref{appendix_simpleswingablation}.

For the inertial movement task, the \textit{Target Encirclement} task, the graph illustrates the change in average distance to the target. The Event-VLC system has approached the most ideal position relative to the target.
Penalties were imposed for collisions, so when agents gather around the target, failure to coordinate with the surroundings would lead to collisions and decreases in reward. VLC could effectively utilize information from surrounding agents through communication. Similarly, for the \textit{Goal Crossing} task, there was a tendency to reach the goal earlier when using VLC.
In tasks where movements involve inertia, it is conceivable that more efficient coordinated actions, such as collision avoidance, can be achieved by visually recognizing nearby agents and obtaining linked communication information.

Additionally, as indicated in Table~\ref{appendixtable} in the Appendix~\ref{appendix_radioid}, in an environment where agents can always obtain information about other agents through radio communication and be identified, the highest performance is achieved. This indicates that factors such as the inability to distinguish opponents and lack of constant access to communication information are elements that degrade performance. As Fig.~\ref{fig:simulationresult} shows, the performance degradation due to the lack of individual identification and the inability to link with communication information is significant. In this chapter, it was demonstrated that depending on the task setup, performance can be enhanced with event-VLC approaches.

%

%% file: 5_results.tex
\begin{figure*}[t]
  \centering
  \includegraphics[width=0.95\linewidth]{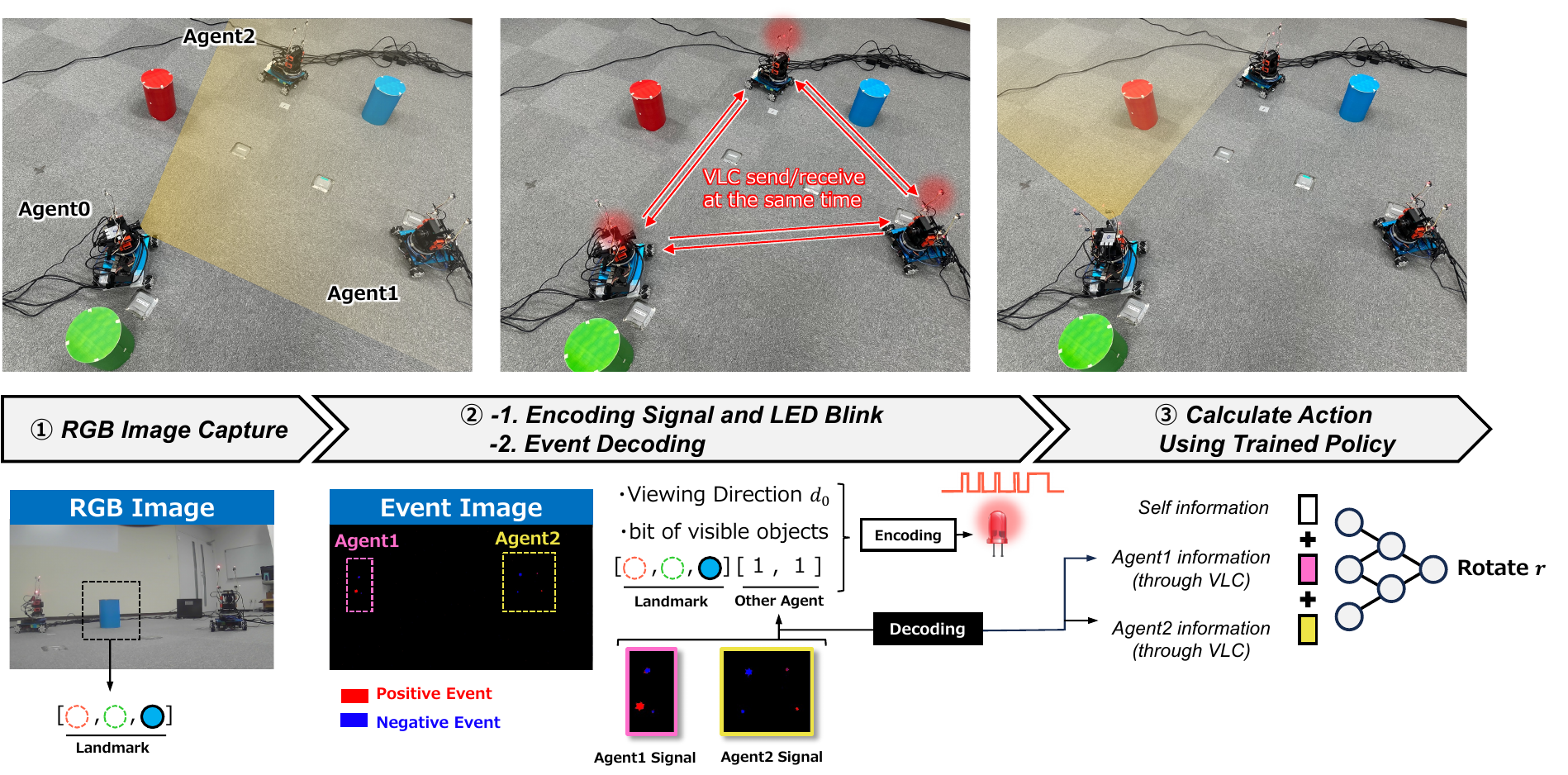}
  \caption{\textcolor{black}{Multi-agent operation using real robots. This shows communication between agents using event-VLC and cooperative behavior using them.}
  }
  \label{fig:experimentconcpt}
\end{figure*}

\section{Experiments with real robots}



\subsection{Settings}

The real-world experimental setup used the physical robot agents described in Section~\ref{approach_actualrobotsingle}. Here, the \textit{Simple Swing} task explained in Section~\ref{4_sim} is performed. For the field setup, landmarks were placed randomly on the edges of an equilateral triangle, similar to the situation described in Section~\ref{4_sim}. The landmarks comprised three objects of red, green, and blue colors. The detection of landmarks was performed using OpenCV. Each agent performed VLC using an event camera, and multiple agents cooperated to ensure no landmarks were missed.

The multi-agent operation of one step is illustrated in Fig.~\ref{fig:experimentconcpt}. Agents determined which landmarks were within their field of view using RGB cameras. They encoded the visual direction and the visible landmark and agent into binary signals, transmitted as flashes of LED lights. Each agent decoded these light signals using event cameras and combined the received information with its own and other agents' information from the past three frames. Using the trained network, agents selected their next visual direction.
This system was designed using Robot Operating System (ROS), and the ROS node relationships are depicted in the Appendix~\ref{appendix_rosnode}.

\subsection{Results}

In the experimental results, the three agents could capture three landmarks in cooperation within an average of $3.22\pm1.54$ steps. 
Furthermore, the communication success rate was 99.1\%. Previous reports employing a photodiode as a receiver for optical wireless communication in mobile systems have indicated a decrease in communication success rate because of the tilt of agents and variability of devices in practical experiments~\cite{nakagawa2023}. In contrast, the communication range of an event camera was broader, and the robustness to ambient light was higher, as shown in Table~\ref{comparevlc}. This high success rate can be attributed to these factors. Information about which entities are visible is transmitted as a bit sequence, reducing the communication volume and contributing to the high success rate. In addition, when using VLC with photodiodes in a multi-agent system, it is necessary to separate the emission timings of LEDs in the temporal domain to prevent signal interference. Therefore, in a system with $N$ agents, $N$ sets of LED emission timings need to be established. However, in the event-VLC experiment conducted in this study, multiple communication signals could be simultaneously decoded, enabling fast operation without the need to separate LED emission timings. This confirms the applicability of event-VLC to larger multi-agent systems.

%% file: 6_conclusion.tex
\section{Conclusion}

In the present study, we investigated the application of camera-based visible light communication using event cameras to multi-agent systems. 
Our findings indicated that visible light communication using event cameras demonstrated high applicability with multi-agent systems in terms of communication range, robustness, and signal separation. 
Furthermore, we constructed a standalone visible light communication system using event cameras and verified its superiority in aspects such as distance measurement, occlusion tolerance, and blur because of movement when compared with existing RGB camera-based ArUco marker recognition methods. We also utilized multi-agent simulations to demonstrate that communication methods such as event-VLC, which link visual information with communication data when agents cannot be visually identified, perform comparable or even outperform the standard radio communication methods.
Finally, in real-world experiments, we showcased the coordinated actions of multi-agents equipped with event cameras and robustly identifiable LEDs for the first time. We verified that this method achieved high communication success rates compared with existing methods using photodiodes.

For future work, we plan to explore methods to accelerate operation speeds, which are currently limited by RGB camera processing, by employing event cameras for environmental observation. We also intend to investigate more compact and precise approaches, such as using hybrid event cameras equipped with both RGB and event pixels.

\balance

%% file: 8-Acknowledgements.tex
\section*{Acknowledgements}
This work was supported by Sony Semiconductor Solutions Corporation. We would like to express our gratitude to the members of the company department who cooperated in the production of the actual machine experiments conducted.

%% file: 7_appendix.tex
\newpage
\appendix
\section*{Appendices}
\addcontentsline{toc}{section}{Appendices}
\renewcommand{\thesubsection}{\Alph{subsection}}

\subsection{The One-Dimensional Visual Information and Basic Processing Network} \label{appendix_1dobs}

\begin{figure}[H]
\begin{center}
  \centering
  \includegraphics[width=0.95\linewidth]{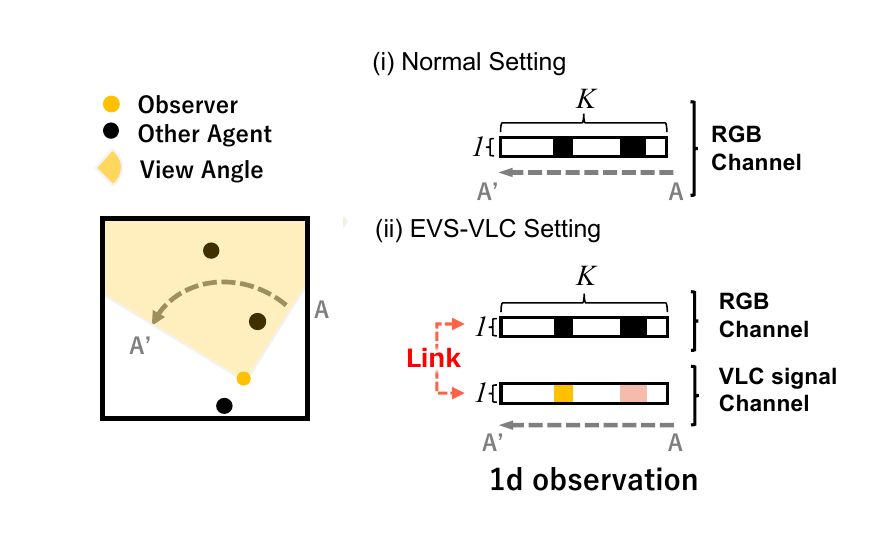}
  \caption{Definition of 1d observation in simulation.
  }
  \label{fig:1dobs}
  \end{center}
\end{figure}

In the simulation, we utilized one-dimensional visual information Fig.~\ref{fig:1dobs}.
In the event-VLC approach, it is assumed that communication signals via VLC can be obtained in the same field of view as RGB information.  In cases where the Field of View (FOV) is less than or equal to 120$^{\circ}$, the parameter $K$ is set to 90. Conversely, when the FOV is 360$^{\circ}$, the parameter $K$ is set to 360.
The field size was set as a square measuring 80 cm on each side, with agents and landmarks sized at 6 cm. 

The basic learning model uses the Actor-Critic method. In the critic component, centralized training is performed using the one-dimensional visual information from all agents.
The individual one-dimensional visual information is combined through a one-dimensional convolutional network to output Q-values. As for the actor component, decentralized execution is carried out using individual information and communication data only. The one-dimensional convolutional network of actor and critic is composed of four-layer 1D-convolutional with 64 filters and a kernel size of 4. The MLP layers are composed of a five-layer MLP with 256 units. In addition, the $f$ and $g$ networks are constructed with four-layer MLP with 256 units.

\begin{figure}[H]
\begin{center}
  \centering
  \includegraphics[width=1.00\linewidth]{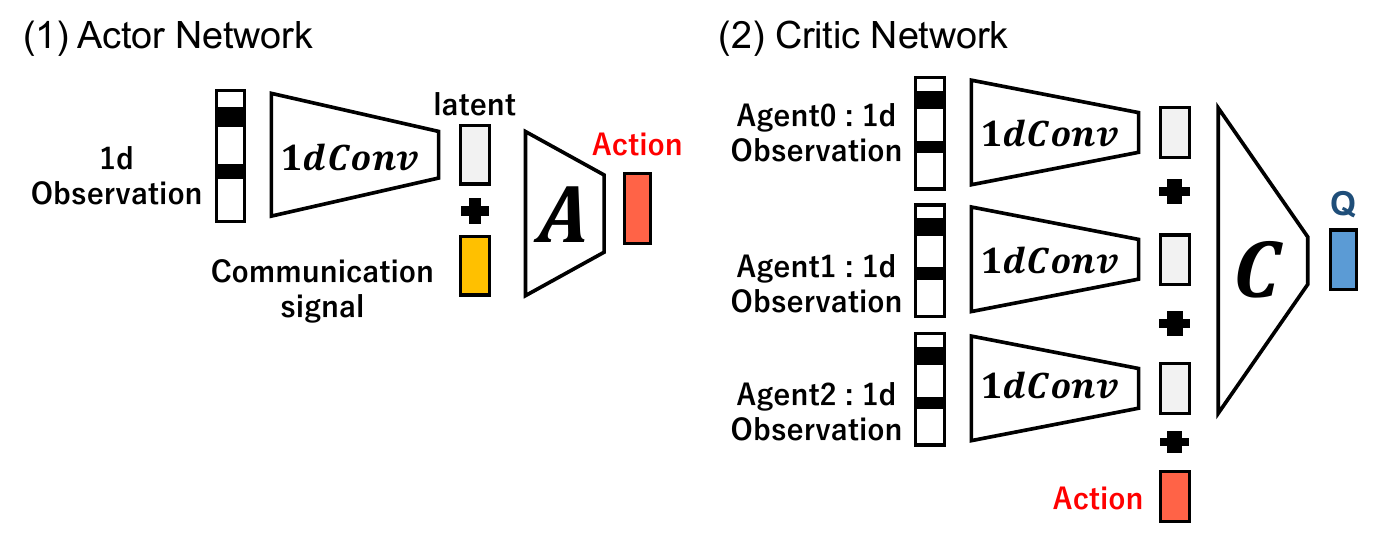}
  \caption{Actor and Critic model.
  }
  \label{fig:testest}
  \end{center}
\end{figure}

\subsection{Model for \textit{Simple Swing}} \label{appendix_simpleswingmodel}


For \textit{Simple Swing} task, which does not involve movement and involve coordination of visual directions, we introduced a network model into the critic that predicts what will be seen when the visual direction is changed.
In our approach, the network $f$ predicts the next topological information and combines its output to compute Q values in the network $C$. During the training process, the network $f$ is trained using the prediction loss of the next topological information, while the network $C$ is trained using the TD loss as its loss function.

\begin{figure}[H]
\begin{center}
  \centering
  \includegraphics[width=0.95\linewidth]{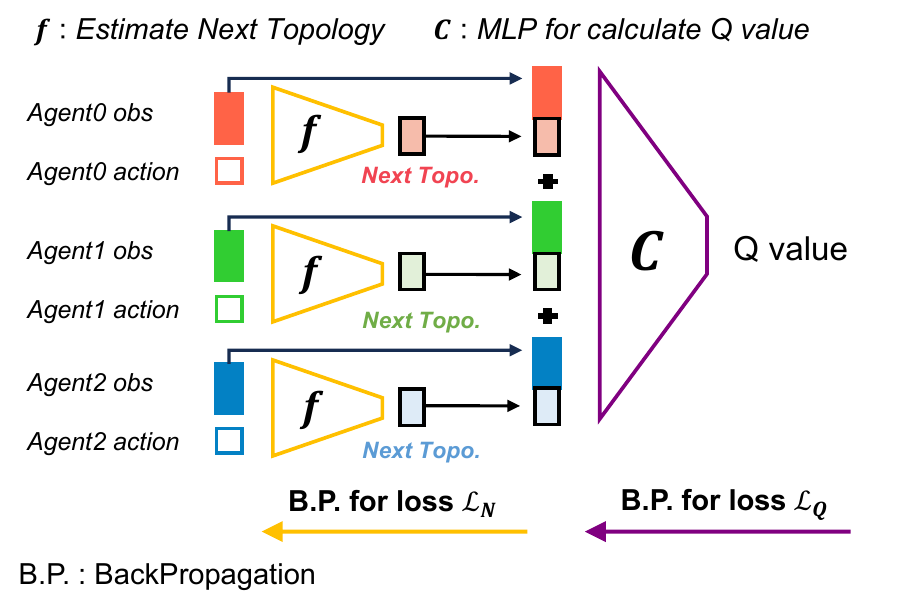}
  \caption{Critic network model of \textit{Simple Swing} task.
  }
  \label{fig:testest}
  \end{center}
\end{figure}

\subsection{Basic Reinforcement Learning Algorithm} \label{appendix_preliminary}

In this study, we consider the behavior of multi-agent as a Markov decision process (MDP).
For a game of $N$ agents, we describe the state of all agents as $S$.
We also denote the observed value of each agent as $O_i$, and each agent takes an action $A_i$.

Letting $\pi_{\phi_i}$ be the measure for the agent to decide on an action, $\phi$ is a parameter of the neural network.
This policy is defined by $\pi:O_i \times A_i$.
Each agent's action acts on the environment and $S$ moves on to the next $S$.
This is the state transition function $T:S \times \sum A_i$ when there are $N$ agents' actions.
Each agent is then given a reward $r_i$ when it takes action $A_i$ in state $S$.
The discount sum $R_i$ is defined as follows using the discount rate $\gamma$.
\begin{equation}
  R_i = \sum \gamma^{\top} r^{\top}_i.
\end{equation}

One method of reinforcement learning is the policy gradient method.
These optimize the agent's parameters $\phi$ so as to maximize the objective function $J(\phi)$.
The gradient of this objective function can be derived as follows.
\begin{equation}
  \nabla_\phi J(\phi) = \mathbb{E}\left[\nabla_\phi \log \pi_\phi (a|s)Q^\pi (s,a)\right].
\end{equation}
Q value in the equation can be calculated by various methods.
In this paper, we use the Actor-Critic algorithm, in which the critic learns to predict $Q(s,a)$ by TD error.
Also, the measure gradient method can be extended to deterministic algorithms.
In deterministic policies, the action for a state is determined according to $\mu_\phi$.
When deterministic action is adopted, the gradient of the objective function can be derived as follows.
\begin{equation}
  \nabla_\phi J(\phi) = \mathbb{E}\left[\nabla_\phi \mu_\phi (a|s)\nabla_a Q^\mu (s,a)|_{a=\mu_\phi (s)}\right].
\end{equation}

In this paper, deterministic policies are employed and the basic learning algorithm is same as MADDPG.

\begin{table*}[]
\renewcommand{\arraystretch}{1.3}
\centering
\caption{Benchmark results including radio with visual identification settings}
\label{appendixtable}
\begin{tabular}{lccccc}
\hline
\multicolumn{1}{c}{\textit{\textbf{Task Name}}} & \textit{\textbf{\begin{tabular}[c]{@{}c@{}}Simple Spread\\ \textnormal{conquest rate {[}$\uparrow${]}}\end{tabular}}} & \textit{\textbf{\begin{tabular}[c]{@{}c@{}}Predator-Prey\\ \textnormal{steps to capture {[}{}$\downarrow${]}}\end{tabular}}} & \textit{\textbf{\begin{tabular}[c]{@{}c@{}}Simple Swing\\ \textnormal{rate of landmarks in sight {[}{}$\uparrow${]}}\end{tabular}}} & \textit{\textbf{\begin{tabular}[c]{@{}c@{}}Target Encirclement\\ \textnormal{distance to the target {[}{}$\downarrow${]}}\end{tabular}}} & \textit{\textbf{\begin{tabular}[c]{@{}c@{}}Goal Crossing\\ \textnormal{arrival rate {[}{}$\uparrow${]}}\end{tabular}}} \\ \hline
event-VLC                                             & 86 ± 0.45 {[}\%{]}                                                                                                    & 8.0 ± 0.34 {[}step{]}                                                                                                      & 97 ± 0.42 {[}\%{]}                                                                                                                & 0.097 ± 0.0016 {[}m{]}                                                                                                                 & 98 ± 0.41 {[}\%{]}                                                                                                      \\
radio                                           & 83 ± 0.50 {[}\%{]}                                                                                                    & 9.5 ± 0.42 {[}step{]}                                                                                                      & 98 ± 0.33 {[}\%{]}                                                                                                                & 0.14 ± 0.0021 {[}m{]}                                                                                                                  & 97 ± 0.50 {[}\%{]}                                                                                                      \\
w/o comm.                                       & 81 ± 0.50 {[}\%{]}                                                                                                    & 12 ± 0.38 {[}step{]}                                                                                                       & 93 ± 0.63 {[}\%{]}                                                                                                                & 0.20 ± 0.0025 {[}m{]}                                                                                                                  & 87 ± 0.85 {[}\%{]}                                                                                                      \\
radio (w/Visual ID)                             & 88 ± 0.58 {[}\%{]}                                                                                                    & 7.8 ± 0.26 {[}step{]}                                                                                                      & 99 ± 0.33 {[}\%{]}                                                                                                                & 0.098 ± 0.0020 {[}m{]}                                                                                                                     & 98 ± 0.39 {[}\%{]}                                                                                                        \\ \hline
\end{tabular}
\end{table*}

\subsection{Ablation Study for \textit{Simple Swing}} \label{appendix_simpleswingablation}

We present an ablation study of the reward setting for the VLC itself and a predictive network model for the next topological information introduced for the \textit{Simple Swing} task. Thus, we can see the impact of increasing overall performance.
From these results, it can be inferred that providing rewards for performing VLC itself and employing a network structure that predicts the next topological information contribute to performance improvement. Providing rewards for VLC itself enables learning of policies that facilitate moderate communication and prevent using only one's own information. 

\begin{figure}[H]
\begin{center}
  \centering
  \includegraphics[width=0.80\linewidth]{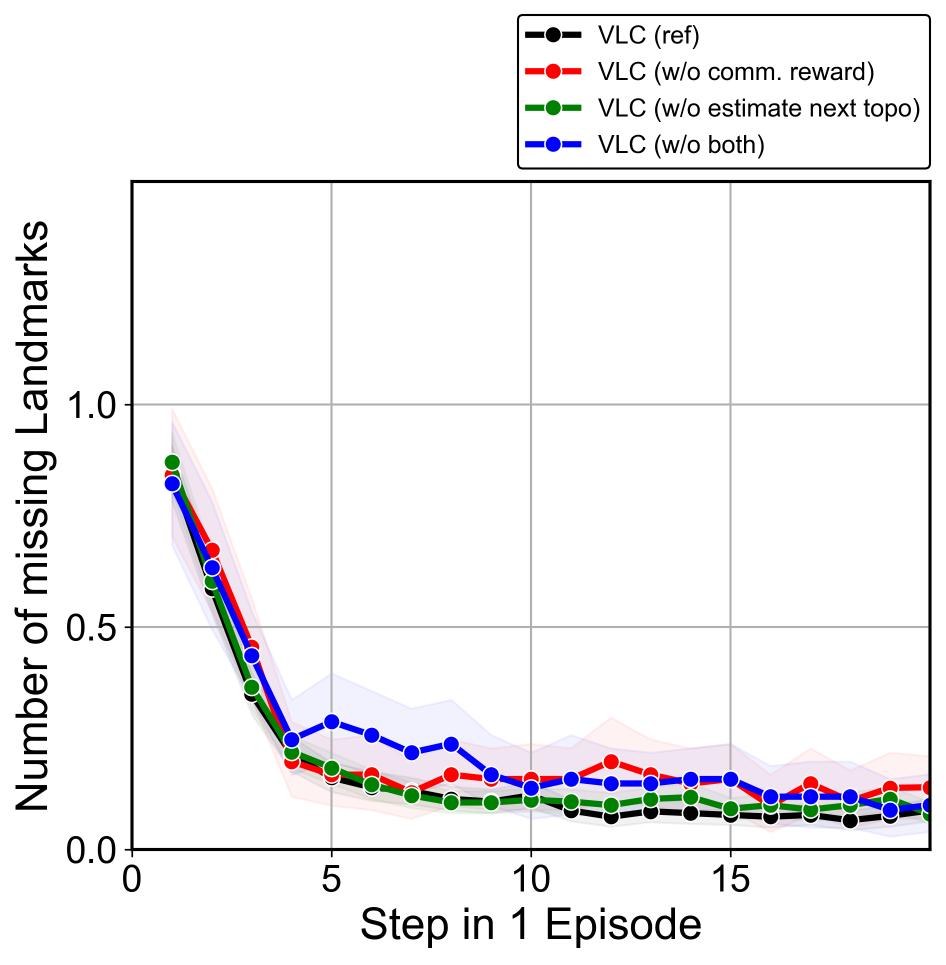}
  \caption{The Ablation Study Results in the Simple Swing Task.
  }
  \label{fig:testest}
  \end{center}
\end{figure}

\subsection{Quantitative benchmark results} \label{appendix_radioid}

\begin{figure*}[]
  \centering
  \begin{minipage}{0.24\hsize}
    \centering
    \includegraphics[width=\linewidth]{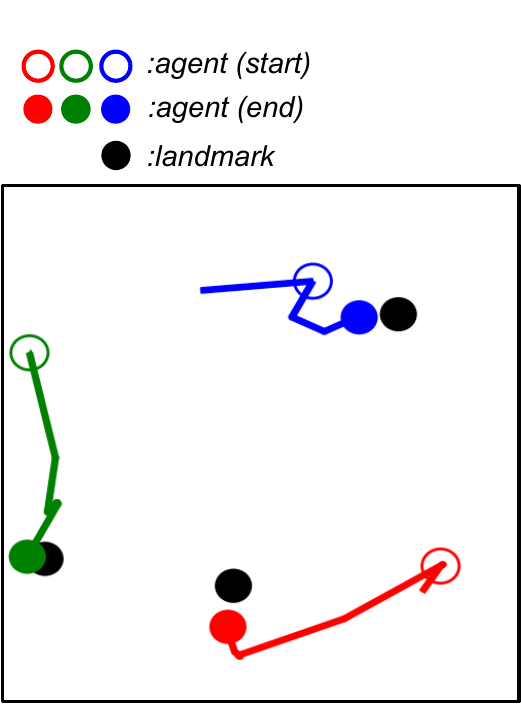}
    \subcaption{\textit{Simple Spread}}
  \end{minipage}
  \begin{minipage}{0.24\hsize}
    \centering
    \includegraphics[width=\linewidth]{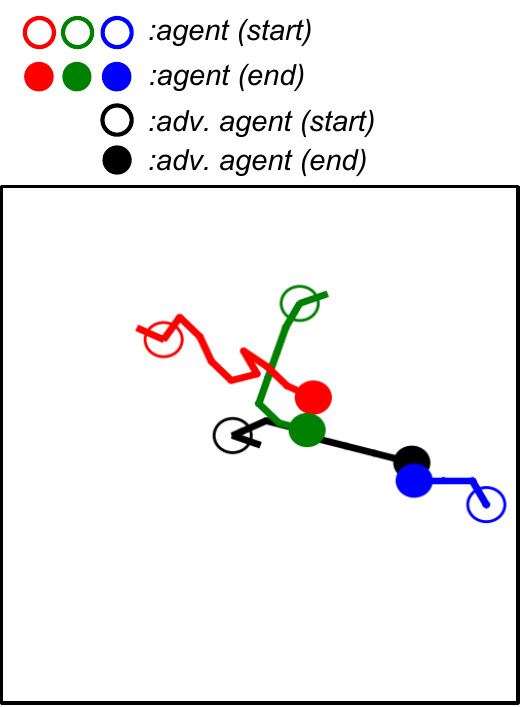}
    \subcaption{\textit{Predator-Prey}}
  \end{minipage}
  \begin{minipage}{0.24\hsize}
    \centering
    \includegraphics[width=\linewidth]{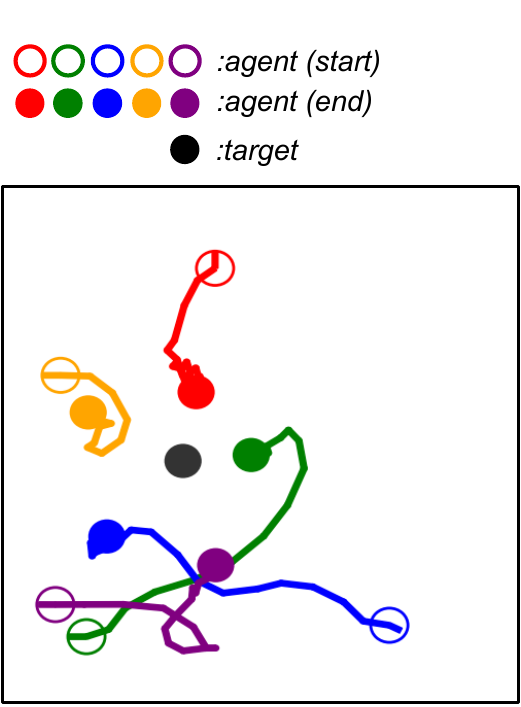}
    \subcaption{\textit{Target Encirclement}}
  \end{minipage}
  \begin{minipage}{0.24\hsize}
    \centering
    \includegraphics[width=\linewidth]{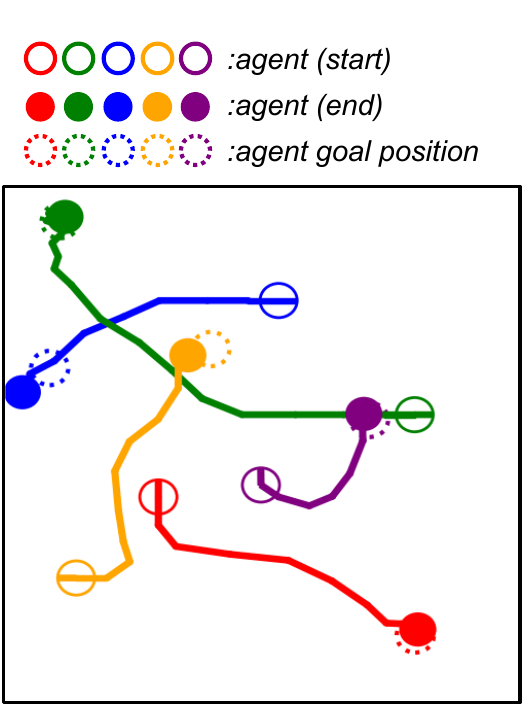}
    \subcaption{\textit{Goal Crossing}}
  \end{minipage}
  
  \caption{
    The trajectories of the learned agents. The agents exhibit optimal behavior in each task.
  }
  \label{fig:trajimage}
\end{figure*}

The results of Fig.~\ref{fig:simulationresult} are summarized quantitatively in Table \ref{appendixtable}. As can be seen, the VLC method performs better in settings where the appearance of other agents is indistinguishable than in the case of radio communication or no communication. On the other hand, it performs best when radio communication plus other agents are distinguishable.
Furthermore, Fig.~\ref{fig:trajimage} illustrating the trajectories of agents for tasks involving movement in \textit{Simple Spread}, \textit{Predator-Prey}, \textit{Target Encirclement} and \textit{Goal Crossing}.
Fig.~\ref{fig:trajimage} shows the behavior of the agents trained in this simulation for tasks involving movement. In each task, the agents are able to cooperate with each other to accomplish the task.

\subsection{The Relationship of ROS Nodes in Multi-Agent Experiments} \label{appendix_rosnode}

\begin{figure*}[]
\begin{center}
  \centering
  \includegraphics[width=0.90\linewidth]{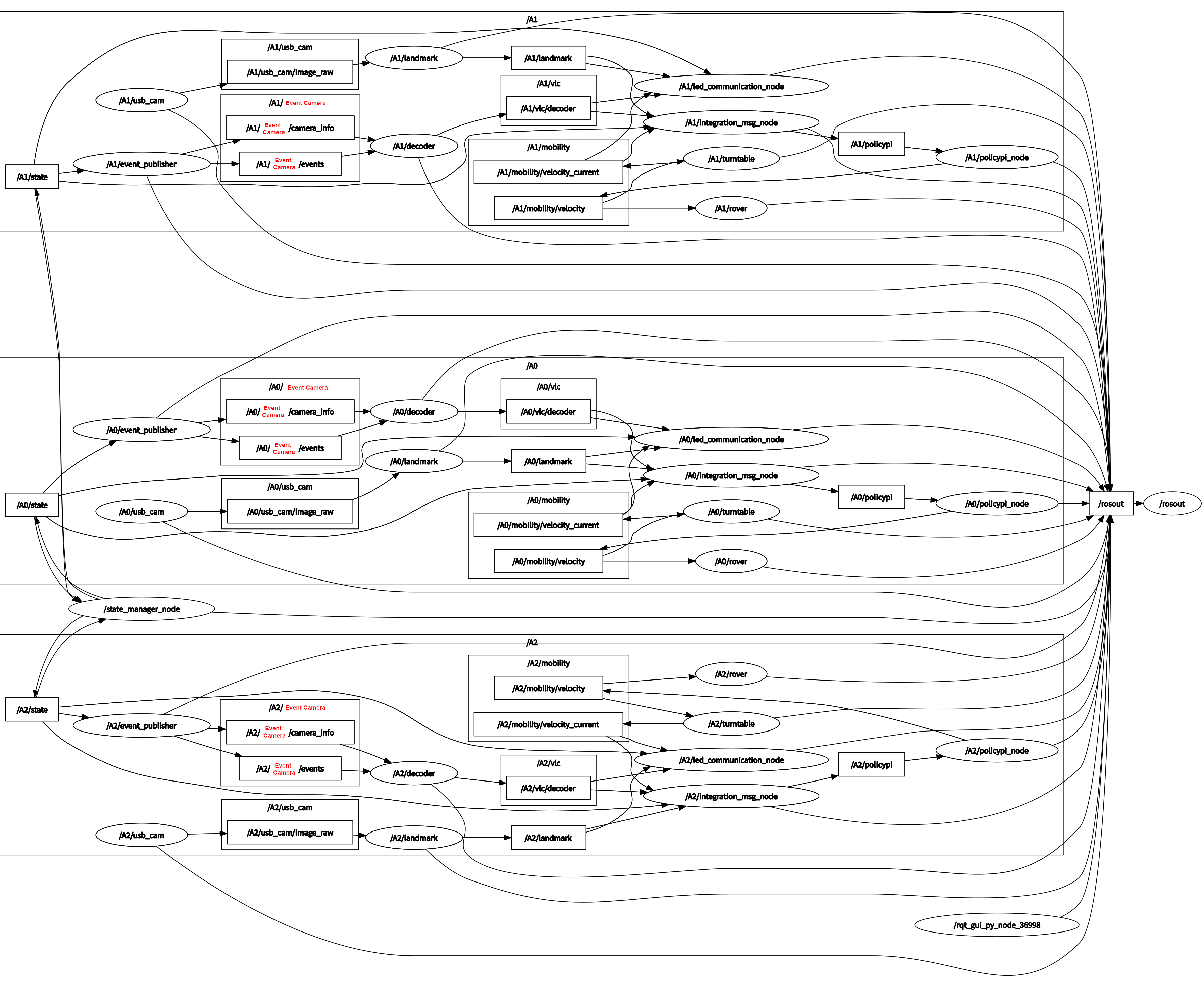}
  \caption{Relationship between ROS nodes.
  }
  \label{fig:rosrelation}
  \end{center}
\end{figure*}

Fig.~\ref{fig:rosrelation} illustrates the relationships of each ROS node in three agents. RGB camera and Event camera acquire information through \texttt{usb\_cam} and \texttt{event\_publisher}, respectively. The images captured by the RGB camera are processed by the \texttt{landmark} node, which maintains information about visible landmarks. Regarding Event information, processed data from the \texttt{decoder} node and the angle of the mobility node's turntable are sent to the \texttt{led\_communication\_node}, which emits signals from LEDs. Similarly, the data is sent to the \texttt{integration\_msg\_node} to calculate the next rotation angle in the \texttt{policypi} node. Certain nodes contained specific terms related to the experiment, which have been revised in red to generic terms.


%
%